\documentclass[prb,aps,amsmath,amssymb,altaffilletter,onecolumn]{revtex4}
\usepackage{epsfig}

\newcommand{\beq}{\begin{equation}}
\newcommand{\enq}{\end{equation}}

\begin{document}

\title{Sequential resonant tunneling in quantum cascade lasers}
\author{Romain Terazzi}
\email{terazzir@phys.ethz.ch}
\author{Tobias Gresch}
\author{Andreas Wittmann}
\author{J\'er\^ome Faist}
\email{jerome.faist@phys.ethz.ch}\affiliation{Quantum Optoelectronics Group,
Institute for Quantum Electronics, ETH, 8086 Z\"urich, Switzerland}

\begin{abstract}
A model of sequential resonant tunneling transport between two-dimensional
subbands that takes into account explicitly elastic scattering is
investigated. It is compared to transport measurements performed on
quantum cascade lasers where resonant tunneling processes are known
to be dominating. Excellent agreement is found between experiment
and theory over a large range of current, temperature and device
structures.
\end{abstract}

\pacs{05.60.Gg, 52.25.Fi, 85.30.De, 42.55.Px }

\maketitle

Resonant tunneling in semiconductor heterostructures has motivated a
large number of experimental and theoretical studies.
One of the most studied case is the resonant tunneling diode,
where a quantum well is formed by a double-barrier region\cite{Gueret:JAP:90:900}.
Under a variable applied electric field, the transparency of the system
is probed by coherent tunneling of electrons at the Fermi energy
of the contact region\cite{Ihn:PRB:96:2315, Eisenstein:PRB:1991}.
The voltage-current
curve exhibits a clear maximum when the emitter electrode aligns
with a resonance of the well.
This perfect case might be realised when the exit barrier
is made so thin that the escape tunneling rate is much faster
than other dephasing mechanisms.
Because of strong in-plane scattering this condition is difficult
to achieve. Usually current proceed by sequential tunneling 
as it is the case in quantum cascade lasers.

In the pioneer work of Kazarinov, the current is expressed in a density matrix
model where the resonance curve is found lorentzian with
an homogeneous broadening given by the average
value of elastic scattering matrix elements. As a consequence
of the averaging the electrons tunnel between
subbands conserving their in-plane wavevectors.

More recently a refined model that include
previously averaged-out second-order
mechanisms was developed\cite{Willenberg:PRB:03:085315-1}.
Second order scattering is known to yield gain without a net
population inversion\cite{Terazzi:NP:07:329} through scattering
assisted optical transitions, but it
also affects more generally resonant tunneling,
by allowing transitions between subband states of different
wavevectors\cite{Willenberg:PRB:03:085315-1}.
It is found that resonant tunneling occurs with conservation of
the energy rather than the wavevector, contrarily to the first order
case\cite{Wacker:ASSP:01}.

In this letter we demonstrate the important role played
by second order mechanisms in sequential resonant
tunneling and therefore in the carrier transport of
semiconductor heterostructures.

When second order terms are considered in the calculation, the
current density between a pair of subbands coupled through
a barrier is expressed as\cite{Willenberg:PRB:03:085315-1}:

\begin{equation}
	\label{soexpr}
	\frac{j}{d} = e\Omega^2\sum_k
	\frac{\gamma_k^1\left(f_k^{2}-f_{q_+}^{1}\right) + \gamma_k^2\left(f_{q_-}^{2}-f_{k}^{1}\right)}
	{\Delta^2 + \left(\gamma_k^1 + \gamma_k^2\right)^2}
\end{equation}

with $q_\pm=\hbar\sqrt{2m^*(\epsilon_k\pm\Delta)}$, where $f_k^i$
is the carrier distribution in subband $i$ at wavevector $k$, $\Delta$
is the detuning between the subband edges $\Delta=\epsilon_2-\epsilon_1$,
$\hbar\Omega$ is the coupling energy through the barrier, $d$ is the
difference between the two centroids of the wavefunctions $d=z_2-z_1$,
$e$ is the elementary charge and $\gamma_k^i$ is the broadening
of state $i$ at wavevector $k$.

When a low density of electrons is distributed thermally in
each subband, with the same electronic temperature $T$ and that
a same and uniform scattering potential is considered
$(\gamma_k^1=\gamma_k^2=\gamma)$, the current
density can be integrated and simply rewritten as:

\begin{align}
	\label{finalexpr}
	\frac{j}{d} = \frac{e\Omega^22\gamma}{\Delta^2 + (2\gamma)^2}\Big\{ \theta(\Delta)&(n_2 - e^{-\beta \hbar|\Delta|}n_1) \\
	\nonumber
	+ \theta(-\Delta)&(e^{-\beta \hbar|\Delta|}n_2 - n_1)\Big\}
\end{align}

where $\theta(x)$ is the Heaviside function, with $\theta(x^-)=0$, $\theta(x^+)=1$ and $\theta(0)=\frac{1}{2}$, $\beta=1/kT$ with $k$ the Boltzmann constant and
$n_i$ is the net population of subband $i$.

\begin{figure}[h]
   \begin{center}
    	\includegraphics[width=85mm]{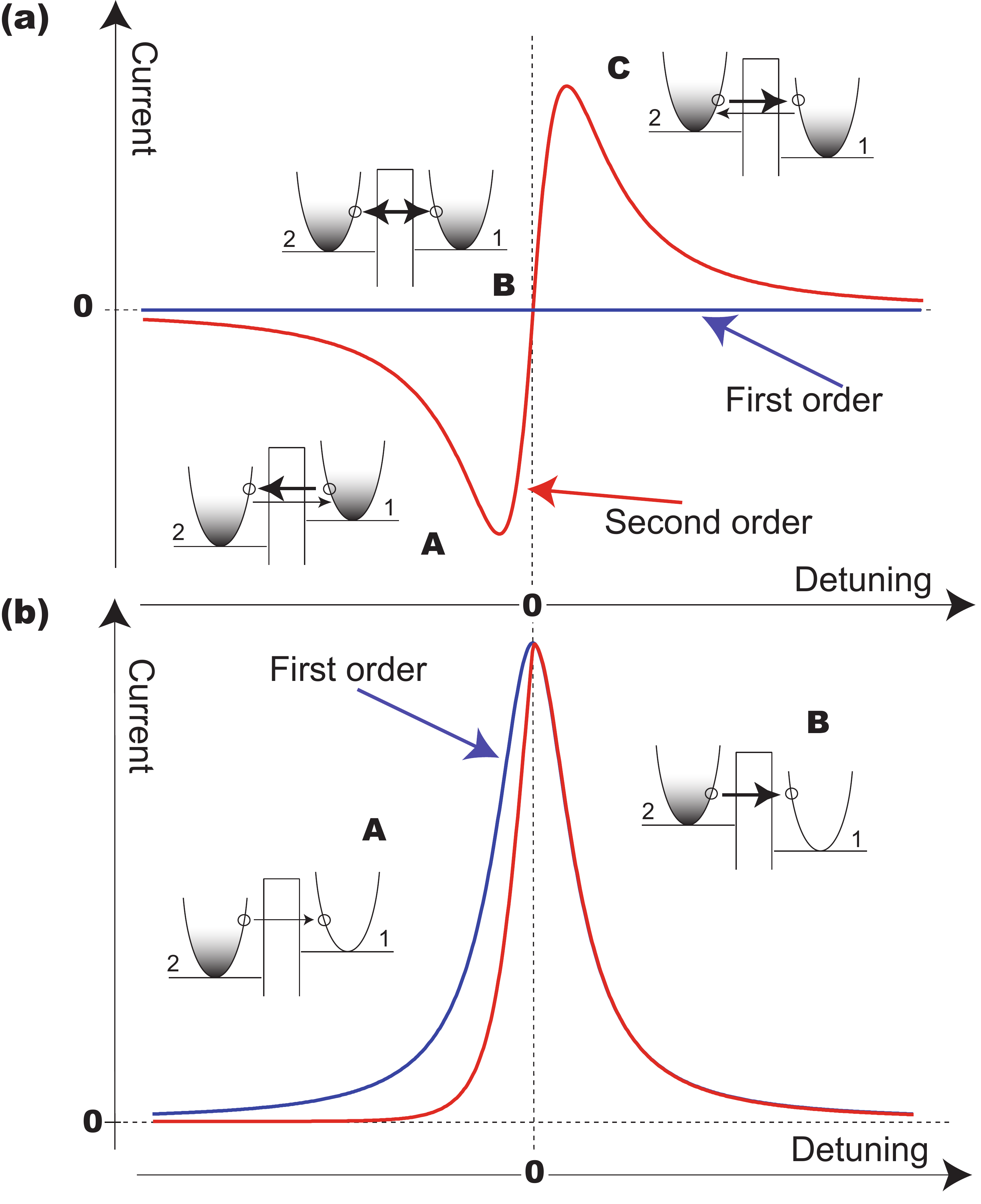}
	\caption{\label{socurrent}(a) Effects of second order contributions on tunneling
	between a pair of equally populated subbands. (A) When the detuning is negative,
	the subband edge of subband 1 is above the edge of subband 2. As tunneling
	conserves energy, the current flow from 1 to 2 is greater than the current
	flow from 2 to 1, yielding a negative net current. (B) When the subbands
	are aligned, the detuning is zero and both contributions cancels, yielding
	a zero net current. (C) The detuning is positive and therefore the edge of
	subband 2 is above the edge of subband 1, yielding a positive net current
	between subbands.
	(b) Empty subband 1. (A) The current is reduced as only a fraction of electrons
	can tunnel. (B) Models overlap perfectly.}
   \end{center}
\end{figure}

The current density is no more driven by the population difference
$n_2-n_1$ but by an effective population term. We want to examine two extreme
case: equally populated subbands $n_2=n_1=n$ and one empty subband $n_1=0$.
The first case is shown in Fig.\ref{socurrent}(a). The current density
is dispersive shaped around the resonance. A
negative current peak occurs
when the detuning is negative, this is when the edge of subband 1 is above
the edge of subband 2. When the subbands are aligned the current
is zero and the first order approximation is recovered.
The current then turns to be positive, after the edge of subband
2 has overcome the edge of subband 1, this is when the detuning is positive.
The dispersive shape is the consequence of electron tunneling
at a constant energy rather than at a constant wavevector. As shown
in Fig.\ref{socurrent}(a) the first order model yield a zero current
for any detuning.
This case illustrate a superlattice: the current is zero until
second-order scattering terms have been taken into account.

The case where one subband is empty is shown in Fig.\ref{socurrent}(b).
For negative detunings the current between the subbands is exponentially reduced
as only the electrons with a sufficient kinetic energy are able to
tunnel to subband 1. Contrarily, for positive detunings, the first and
second order curves overlap perfectly as all electrons are above the
edge of subband 1.

Generally the first order model is recovered as the thermal energy
largely overcomes the detuning energy $kT\gg \hbar|\Delta|$,
as it is the case in\cite{Kazarinov:SPS:72:120}. In Eq.\ref{finalexpr},
the exponential cut-off tends to $1$ as the temperature tends to
infinity, spreading electrons in an uniform distribution.

As second order mechanisms strongly affect the resonant current between 
a pair of subbands, we aim to show its impact on more complex semiconductor
heterostructures like quantum cascade lasers\cite{Faist:Sci:94:553}.
We therefore have implemented second-order effects in the computation
of the voltage-current characteristic.

\begin{figure}[h]
   \begin{center}
    	\includegraphics[width=85mm]{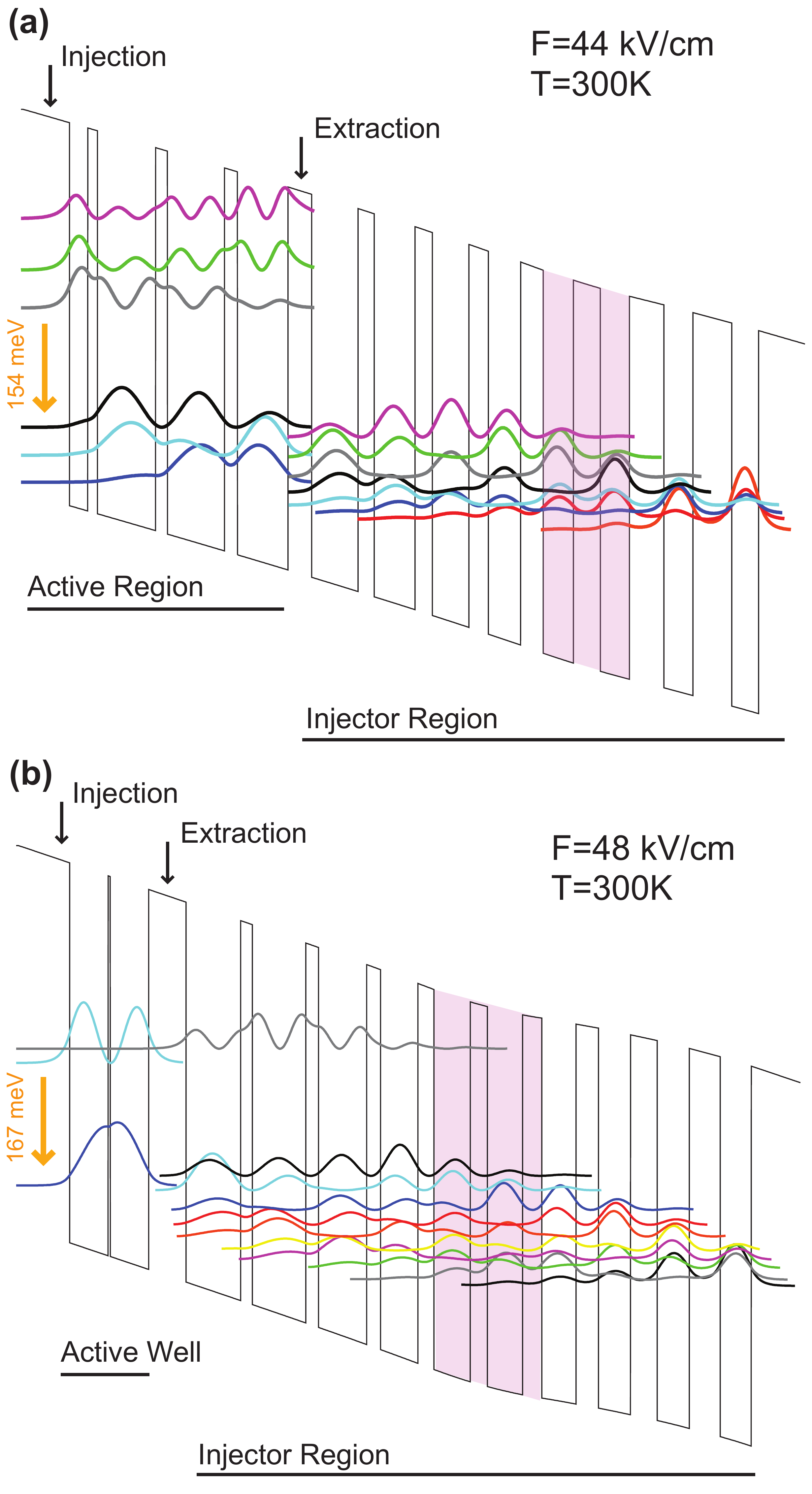}
	\caption{\label{n655-n257}}
	Each structure is shown at injection resonance field. The layer
	sequence starts from the injection barrier and the thicknesses are in nm;
	roman, resp. bold, numbers indicate $\text{In}_{0.53}\text{Ga}_{0.47}\text{As}$, resp.
	 $\text{Al}_{0.48}\text{In}_{0.52}\text{As}$ alloy, acting as well, resp. barrier
	 material.
	(a) Layers: {\bf4.3}/1.7/{\bf0.9}/5.4/{\bf1.1}/5.3/{\bf1.2}/4.7/{\bf2.2}/4.3/{\bf1.5}/3.8/{\bf1.6}/3.4/{\bf1.8}/3.0/{\bf2.1}/
	\underline{2.8}/\underline{{\bf2.5}}/\underline{2.7}/{\bf3.2}/2.7/{\bf3.6}/2.5.
	Underlined layers are $1.5\times10^{17}\;\text{cm}^{-3}$  Si doped.
	Nominal sheet carrier density is $1.2\times 10^{11}\;\text{cm}^{-2}$.
	Period length is 68.3~nm, repeated 35 times.
	The optical transition occurs at $\approx$154~meV.
	(b) Layers:
	 {\bf4.8}/3.6/{\bf0.2}/3.6/{\bf3.5}/5.1/{\bf1.1}/5.0/{\bf1.2}/4.5/{\bf1.3}/3.5/{\bf1.5}/\underline{3.4}/ \underline{{\bf1.6}}/\underline{3.3}/\underline{{\bf1.8}}/
	3.2/{\bf2.1}/3.0/{\bf2.5}/3.0/{\bf2.9}/2.9.
	Underlined layers are $3\times10^{17}\;\text{cm}^{-3}$  Si doped.
	Nominal sheet carrier density is $3.03\times 10^{11}\;\text{cm}^{-2}$.
	Period length is 68.6~nm, repeated 35 times.
	The optical transition occurs at $\approx$167~meV.
   \end{center}
\end{figure}

The computational model is based on the density matrix where dissipation
is included as rate equations for the populations and as dephasing
times for the polarisations. The precise implementation
will be detailed somewhere else\cite{Terazzi:Unpublished:08}.
A typical quantum cascade laser is a
repetition of a fundamental period as shown in Fig.\ref{n655-n257}.
These periods are coupled through
an injection barrier. Electrons are injected by sequential
resonant tunneling 
from a period to the next one. The period itself can be separated
in an active region where the laser transition occurs and an injector
region where carriers are relaxed before they are injected into the
next period.
For many structures, and the ones presented here,
the active region is coupled to the injector region through an extraction
barrier as shown in Fig.\ref{n655-n257}.
We therefore have implemented tight-binding and sequential resonant tunneling
at the injection and at the extraction barrier. In the active region
and the injection region, the carriers are relaxed through intersubband
scattering. The mechanisms we have considered\cite{Unuma:JAP:03:1586,
Tsujino:APL:05:062113-1} include LO-Phonons,
interface roughness and ionized impurities (dopants) scattering.
Non-parabolicity effects and self-consistency of the potential
are accounted by the model. A uniform electronic temperature is
computed for all subbands, based on the electron energy balance\cite{Harrison:JAP:02:6921}.
The numerical simulations output the populations of the subbands and the
current density flowing through the heterostructure.
We are therefore able to predict the voltage-current
characteristic of a particular quantum cascade structure. In order to test
the impact of second order transport, we have implemented
both first and second order sequential resonant tunneling models.

We present two quantum cascade structures in different coupling regimes.
The first show in Fig.\ref{n655-n257}(a) has a strong coupling between the active and the
injector regions as the extraction barrier is made sufficiently thin (22~\AA).
Contrarily in Fig.\ref{n655-n257}(b) the second structure is a
single quantum well as its active region is formed by one well only,
weakly coupled to the injector region by a thick extraction barrier (30~\AA).

\begin{figure}[h]
   \begin{center}
    	\includegraphics[width=85mm]{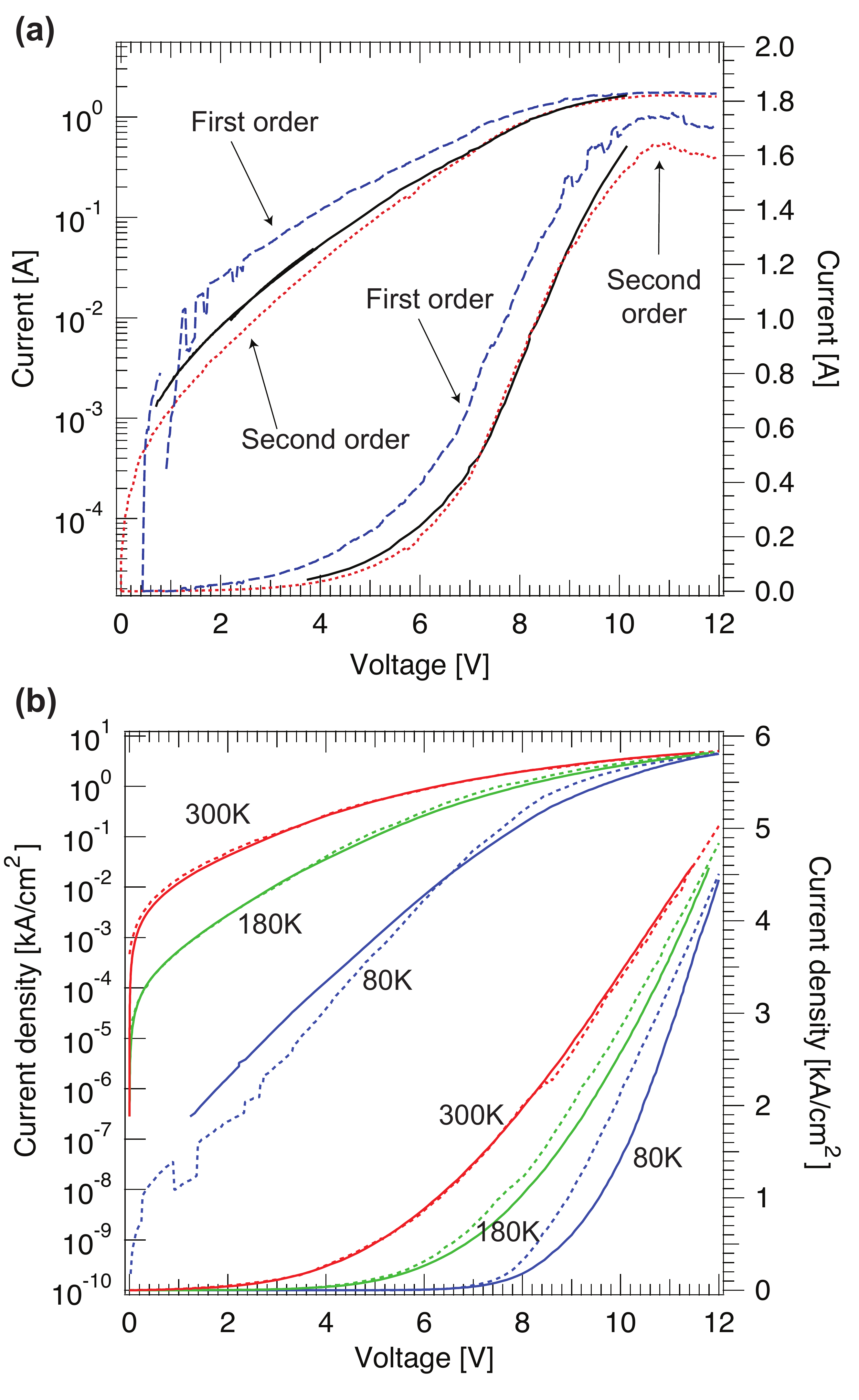}
	\caption{\label{liv}}The current curves are shown both
	in log scale (left axis) for inspection of low currents
	and in linear scale (right axis) for inspection of the dynamic
	range. (a) Experimental data (full line),
	simulation with second order resonant tunneling (dotted line), simulation
	with first order model (dashed line).
	(b) Single quantum  well current-voltage curves for three temperatures:
	80~K, 180~K and 300~K. Measurements displayed in full line. Simulations
	in broken line.
   \end{center}
\end{figure}

The current-voltage curves for both structures are shown in Fig.\ref{liv}.
The measurements are taken in continuous mode for low currents and
in pulsed mode when current flow causes a heating of the sample. The cryostat temperature
for the first sample is 300~K, while it is 80~K, 180~K and 300~K
for the second.

In Fig.\ref{liv}(a) the second order
voltage-current curve fits the experimental data much better than
the first order approximation from the very low currents to the maximal current.

If we focus on the low field values of the voltage-current curve,
the second order model clearly better predicts the experimental
behaviour than the first order model does.
In particular it yields a zero net current at
zero field which is an important validation of the computational
model.

The result for the single quantum well structure are shown
in Fig.\ref{liv}(b). We have shown simulated curves with
second order model only, because the first order model
failed to converge at low fields values and is largely off
from the measurements.
Apart from a constant serial resistance (0.5~$\Omega$) than higher systematically
the experimental bias, the model was able to reproduce nicely the experiment,
in particular for high temperature where the transport
in the structure is clearly dominated by optical phonons.
The model predicts the low temperature curve with less
accuracy because the transport at such a temperature
also require the computation of scattering rates due
to acoustical phonons and electron-electron interactions,
which are not computed in the present model.

Agreement between computed and experimental current-voltage
characteristics has already been reported for other model
approachs such as based on Monte Carlo\cite{Callebaut:JAP:98:104505} or non-equilibrium
Green's functions\cite{Wacker:PRB:02:085326}; the comparison was done however on
a much more limited range of currents, temperature and structure
design.

Formally, the current driven by tunneling between two subbands
through a barrier or by optical absorption are physically equivalent
because both processes conserve the in-plane wavevector.
As a result, the striking agreement between the predictions
of the second-order model and the experiment can be interpreted
as a strong experimental evidence for the validity of the Bloch
gain model.

This work was supported by the Swiss National Science Foundation,
the National Center of Competence in Research, Quantum Photonics
and the Swiss Commission for Technology and Innovation.

\bibliographystyle{apsrev}

\end{document}